\def \yskip{\penalty-50\vskip3pt plus 3pt minus 2pt}
\def \reference{\par \yskip \noindent \hangindent .4in \hangafter 1}
\def \abc#1#2#3#4 {\reference#1, {\sl#2}, {\bf#3}, #4}
\def \blank {\lower 5pt\hbox to 0.75in{\hrulefill}}

\def \cm{~\rm{cm}}
\def \s{~\rm{s}}
\def \km{~\rm{km}}

\def \g{~\rm{g}}
\def \AU{~\rm{AU}}
\def \yrs{~\rm{yrs}}
\def \yr{~\rm{yr}}
\def \K{~\rm{K}}

\def \erg{~\rm{erg}}

\def \lae{\mathrel{<\kern-1.0em\lower0.9ex\hbox{$\sim$}}}
\def \gae{\mathrel{>\kern-1.0em\lower0.9ex\hbox{$\sim$}}}

\documentstyle[11pt,aasms,tighten,flushrt]{article}
\begin{document}
%\normalsize
\small

\setcounter{page}{1}
%\noindent Presented at the 180 IAU Symposium: {\it Planetary Nebulae}, 
%August 1996.
%dust1.tex
\begin{center} \bf 
DUST FORMATION AND INHOMOGENEOUS MASS LOSS 
FROM ASYMPTOTIC GIANT BRANCH STARS
\end{center}
%\vspace*{2.0cm}

\begin{center}
Noam Soker\\
Department of Physics, University of Haifa at Oranim\\
%Mathematics-Physics\\
Oranim, Tivon 36006, ISRAEL \\
soker@physics.technion.ac.il 
\end{center}

%\clearpage 

\begin{center}
\bf ABSTRACT
\end{center}

We examine the flow from asymptotic giant branch (AGB) stars when along
a small solid angle the optical depth due to dust is very large.
 We consider two types of flows.
 In the first, small cool spots are formed on the surface of slowly
rotating AGB stars.
  Large quantities of dust are expected to be formed above the surface of
these cool spots.
  We propose that if the dust formation occurs during the last AGB phase
when mass loss rate is high, the dust shields the region above it
from the stellar radiation.
 This leads to both further dust formation in the shaded region,
and, due to lower temperature and pressure, the convergence of the
stream toward the shaded region,
{{{ and the formation of a flow having a higher density than
its surroundings. This density contrast can be as high as $\sim 4$.
 A concentration of magnetic cool spots toward the equator will lead
to a density contrast of up to a few between the equatorial and polar
directions. }}}
 This process can explain the positive correlation between high mass
loss rate and a larger departure from sphericity
{{{ in progenitors of elliptical planetary nebulae}}}.
 In the second type of flow, the high density in the equatorial plane is
formed by a binary interaction, where the secondary star is close to, but
outside the AGB envelope.
  The shielding of the radiation by dust results in a very slow and
dense flow in the equatorial plane.
 We suggest this flow as an alternative explanation for the equatorial
dense matter found at several hundred astronomical units 
around several post-AGB binary systems.
           
\noindent
%{\it Subject heading:}   % to APJ
{\it Key words:}         % to MNRAS
    Planetary nebulae: general
$-$ stars: AGB and post-AGB
$-$ stars: mass loss
$-$ circumstellar matter

%\clearpage 

% ======================================================================
\section{INTRODUCTION}
% ======================================================================

 The generally accepted model for the high mass loss rate on the upper
asymptotic giant branch (AGB) includes strong stellar pulsations
coupled with large quantities of dust formation at a few stellar
radii around the stellar surface (e.g., Wood 1979; Jura 1986; Knapp 1986; 
Bedijn 1988;  Bowen \& Willson 1991; Fleischer, Gauger \& Sedlmayr 1992;
Habing 1996; H\"ofner \& Dorfi 1997;
{{{ Andersen, Loidl, \& H\"ofner 1999}}}).
 Radiation pressure on dust accelerates the dust particles which
drag the gas particles with them.
 Pulsations help in several ways.
 The main processes are the formation
of an extended envelope by pulsational-driven shocks
(Bowen 1988; Bowen \& Willson 1991), and the cooling behind shock waves
(Woitke, Goeres, \& Sedlmayr 1996).
 Both from theory (Bowen 1988) and observations (Jura 1986; Knapp 1986)
it turns out that mass loss rate is positively correlated with
the pulsation period.

  Cool spots on the photosphere can further enhance local dust formation,
hence local mass loss rate.
 Such cool spots were suggested to occur in cool giants (e.g.,
Schwarzschild 1975; Polyakova 1984; Frank 1995; Soker 1998) and
on the surface of R Coronae Borealis (RCB) stars
(Clayton, Whitney, \& Mattei 1993; Soker \& Clayton 1999).
 In the works of Polyakova (1984), Clayton {\it et al.} (1993),
Soker (1998), and  Soker \& Clayton (1999), the cool spots are
assumed to be due to  magnetic activity.
 If the cool spots are concentrated near the equatorial plane,
as is expected for magnetic cool spots (Soker 1998;
Soker \& Clayton 1999), this will lead to enhanced equatorial mass
loss rate and to the formation of an elliptical planetary nebula (PN).
 The magnetic cool spots model has the advantage that it can explain
the increase in the departure from spherical mass loss on the final stages
of the AGB, and during the early post-AGB phase (Soker \& Harpaz 1999),
and can be effective for very slowly rotating AGB stars,
having rotation velocities as low as $\sim 3 \times 10^{-5}$ times the
equatorial Keplerian velocity (Soker \& Harpaz 1999).
 The departure from sphericity caused by cool magnetic spots
is both in local inhomogeneities, such as filaments and arcs,
and a higher equatorial mass loss rate. 
 Soker \& Harpaz further argue that the changes of the envelope properties
of upper AGB stars caused by a small amount of mass loss are sufficient
to cause large changes in the mass loss geometry, through, e.g., 
enhanced formation of magnetic cool spots near the stellar equator.
 The enhanced dust formation above magnetic cool spots will lead to large
density inhomogeneities in the wind, which will form filaments, arcs, and
loops in the circumstellar envelope, and later in the PN.
  In the present paper we examine whether the density inhomogeneities
can be further increased {\it after} dust formation, but when the material
is still close to the stellar surface.
  A full study of the flow structure requires two-dimensional numerical
simulations, which include dust formation, interaction of dust with gas
and radiation, radiative cooling, heating by the
central star's radiation, shocks generated by the stellar pulsation, as well
as the usual gasdynamical equations.
 The present paper is limited to a preliminary exploration of the
proposed flow structure and the density inhomogeneities above cool spots.

 Based on their high resolution observations
of the red supergiant star VY CMa,  Monnier {\it et al.} (1998)
argue for a locally enhanced dust formation and mass loss.
 This star is more massive $\sim 25 M_\odot$, and more luminous
$\sim 2 \times 10^{5}L_\odot$, and has a more extended envelope
$\sim 8.3 \AU$ than AGB stars.
 Based on the curvature of the dusty plume, Monnier {\it et al.} (1998)
propose that the star rotates at an angular velocity of $\gtrsim 10^{-3}$
times the equatorial Keplerian velocity, larger than the minimum
required by the cool spots model for AGB stars.
 This star shows that a locally enhanced dust formation from
rotating red giants is a reasonable assumption, and can lead to the
formation of a filamentary circumstellar envelope.
 More support for the local formation of dusty clouds close to the surface
of evolved stars comes from R Coronae Borealis stars
(Clayton {\it et al.} 1999 and references therein).

 To conduct the exploratory study we start in $\S 2$ with an outline
of the proposed scenario, and in $\S 3$ we continue by examining
the relevant time-scales.
 In $\S 4$ we discuss possible implications of optically thick dusty flow
on the mass loss process from close binary systems.
 Our discussion and summary are in $\S 5$
                                                       
% ======================================================================
\section{THE FLOW STRUCTURE}
% ======================================================================

  We classify three basic types of flow near the surface of single slowly
rotating stars, i.e., the centrifugal force is negligible.
 For this we define the optical depth in the inner acceleration zone$-$the
zone above the star where the gas accelerates due to pulsations and
radiation pressure on dust.
  According to Bowen (1988), this occurs inside a radius of
$r_u \lesssim 3 R_\ast$, where $R_\ast$, is the average
(over the pulsation period) stellar radius.
 The optical depth in the acceleration zone is therefore
\begin{equation}
\tau_a = \int_{R_\ast}^{r_u} \kappa(r) \rho(r) dr,
\end{equation}
where $\kappa(r)$ is the opacity.

   In what we term the {\it optically thin} flow, the dust and gas
optical depth in the acceleration zone is always much smaller than
unity, i.e., $\tau_a \ll 1$.
 In the {\it locally optically thick} flow, the dust and gas optical
depth above the cool spot $\tau_{\rm sh}$ is much larger than unity,
while its value along other directions $\tau_e$ is much smaller than unity.
 In the {\it optically thick} flow, the dust opacity is always much
larger than unity, for $r_u \lae 3 R_\ast$, and in all directions.
This means that dust forms very close to the stellar surface
on the entire surface.

 The optically thin flow is likely to occur when the mass loss rate is low,
during the AGB before the superwind starts, and during the late post-AGB
phase.
 It depends on when cool spot activity becomes strong, which in turns
depends on the envelope properties (Soker \& Harpaz 1999) and on the
envelope's angular velocity. 
 If the optical depth is low, the material above the cool spot will be
only slightly cooler than the surroundings, by a factor of $<1.5$
(Frank 1995).
 If pressure equilibrium between the shaded region$-$the region above the
cool spot$-$and the surroundings is achieved, as is likely to be the case,
the density contrast will not be large.
 Even if initially more gas is expelled above the spot, its larger thermal
pressure means that it will expand in the transverse direction,
and the density contrast between the shaded region and
surroundings may even {\it decrease}.

  Since the optically thick flow requires a cool photosphere, it may
occur in extended AGB stars which lose mass at a very high rate.
 Many, and probably most, stars on the upper AGB do have a dusty optically
thick circumstellar envelope, but most of the opacity is due to dusty
material farther away than $r_u \sim 2 R_\ast$.
 Since this type of flow is rare and seems to be very complicated,
we do not consider it in this preliminary study.

 We concentrate therefore on the {\it locally optically thick} flow
above cool spots.
A schematic presentation of the dust distribution and flow structure is
plotted in Figure 1.
 Dust is assumed to form in large quantities very close to the cool spot's
surface.
 It shields the region behind it from radiation$-$the shaded region.
  Temperature in the shaded region is much below its equilibrium value,
more dust forms, and its grain sizes increase.
  At about $\sim 2 R_\ast$ dust is formed around the entire star as well,
but its density is lower, and optical depth is below 1. 
 The higher temperature of the surroundings means higher pressure, hence
material streams into the shaded region, forming more dust.
 The result is that the stream near the symmetry axis of the conical
shaded region is not radial but cylindrical.
  Therefore, away from the star, where the expansion velocity is
constant, the density will drop less steeply than $r^{-2}$, as long as the
temperature of the surroundings is higher than that of the shaded region.

 An interesting result of the dust shielding is the required spot size. 
Without shielding, the temperature above a cool spot does not fall 
with radial distance from the surface as steeply as the temperature 
of the environment.
 This is so because as we move away from the spot on a radial direction 
we see more of the normal photosphere, and less of the cool spot 
(Frank 1995). 
 For the region to stay cool enough to form dust, the spot must be large: 
its radius should be $b_s \gtrsim 0.5 R_\ast$ (Frank 1995). 
 This is a large spot, which is not easy to form by concentration of
small magnetic flux tubes (Soker \& Clayton 1999). 
 However, with the dust forming very close to the surface, as is suggested
for cool magnetic spots (Soker \& Clayton 1999), the shielded region forms
dust, which in turn shields a region farther away. 
 Therefore, when dust forms very close to the surface of a cool spot with
high optical depth, dust will be formed in the entire shaded region 
even when the spots is much smaller than what is required without dust
shielding.
  Not only does the proposed flow allow higher mass loss rate from small
cool spots, it is also limited to small spots.
 Above a large spot the relative mass entering from the surroundings is
small, and since the radiation from the spot is weaker, the material
in the shadow will not be accelerated much.
 Large cool spots can lead to a higher mass loss rate in the optically
thin flow, where they allow both dust formation closer to the surface
(Frank 1995) and radiation to accelerate the flow.

 Until now we have considered a steady state flow. 
However, the formation and evolution of a cool spot, and the process of dust 
formation coupled with the stellar pulsations, will lead to sporadic 
behavior  (e.g., for spherical dust acceleration see
H\"ofner, Feuchtinger, \& Dorfi 1995 and
{{{ Fleischer, Gauger \& Sedlmayr 1995}}}).
 In particular, dust may be formed in large quantities; it blocks the
radiation, and the material farther out, instead of being accelerated
outward, will be pulled back to the surface by the stellar gravity,
before the flow depicted in Figure 1 reaches its
``steady state'' behavior.
 The average mass loss rate above the spot is still higher because of dust 
formation, but a temporary and locally back-flow of material still located
close to the surface, at a distance of less than a few stellar radii, 
is possible. 
 In any case, the outflow velocity of the matter in the shaded region is
slower than in other directions due to the weaker radiation from the cool
spot. 

% ======================================================================
\section{RELEVANT PROCESSES AND TIME SCALES}
% ======================================================================

  We assume that large quantities of dust form above a cool spot
of size $2 b_s$ (see fig. 1), and that the dust blocks most of the stellar
radiation.
 Following the discussion in the previous section we consider a small spot
with a radius of $b_s < 0.3 R_\ast$, so that the area covered by the
spot is smaller by more than an order of magnitude than the stellar disk
area.
 The opacity of dust and gas around AGB stars is
$\kappa \sim 20 \cm^2 \g^{-1}$
(e.g., Jura 1986 for a wavelength of $2 \mu$m).
 The density just above the photosphere of an AGB star of radius
$\sim 2 \AU$ and mass loss rate of $\sim 10^{-5} M_\odot \yr^{-1}$ is
$\gtrsim 5 \times 10^{-14} \g \cm^{-3}$. 
 Using these values we find the optical depth through the gas and
dust along a radial distance $dr$ close to the surface to be
$\tau \sim 10 (dr/10^{13} \cm)$.
 Therefore, when the dust forms very close to the star, the radiation will
be blocked along the radial direction.
 The dust re-emits and scatters the radiation, but in all directions, 
and there will be a shadow behind the region where dust forms.
 Even when the optical depth is not much above unity, and some radiation
penetrates, the radiation is still very weak since the luminosity per unit
surface from a cool spot is much lower than that of the rest of the star.  
 The situation is different when the region above the entire star,
or most of it, has a high optical depth, since the radiation must 
escape eventually.
 There will be no real shadow, as optically thick regions in other
directions will scatter and re-emit radiation in the considered direction. 

 The region behind the dust is in a shadow, and it will cool faster than
its environment.
 Since we require that the cooling time be much shorter than the flow time,
we also neglect the adiabatic cooling which proceeds at the same rate as
the flow time.
 Also, the shaded region will be compressed by the surroundings, reducing
somewhat the efficiency of the adiabatic cooling.
  The cooling time of gas at the relevant temperature range 
($100 K \lesssim T \lesssim 1,500 \K$) and density range
($\rho \sim 10^{-14} \g \cm^{-3}$) is
\begin{equation}
t_{\rm cool} \simeq \frac{3}{2} \frac {kT}{\chi} \simeq
1 \left( \frac{T}{1,500 \K} \right)
\left( \frac{\chi}{10^{-20} \erg \s^{-1}} \right)^{-1}
\yrs ,
\end{equation}
where $\chi=\Lambda/n(H_2)$ is taken from figure 3 of
Neufeld, Lepp \& Melnick (1995); $\Lambda$ is the cooling rate per unit
volume, and $n(H_2)$ is the molecular hydrogen density.
 We assume that the wind above a cool spot is mainly in the form of $H_2$.
 The flow time along a radial distance $l$ is given by 
\begin{equation}
t_{\rm flow} \simeq 
1 \left ( \frac{l}{2 \AU } \right)
\left( \frac{v_f}{10 \km \s^{-1}} \right)
\yrs,
\end{equation}
where $v_f$ is the flow velocity. 

 Examining figure 3 of Neufeld {\it et al.} (1995) we see that the
cooling time is shorter than the flow time when $T \gtrsim 300 \K$. 
 For the temperature of the surrounding gas to be much higher,
say $T_e \sim 1,200 \K$, the flow depicted in figure 1 must occur close 
to the star.
 From the equilibrium temperature around AGB stars (e.g.,
Bowen 1988 fig. 3; Frank 1995 fig. 4) this means that the flow depicted
in Figure 1 occurs at $r \lesssim 3 R_\ast$. 
 At $r \lesssim 2 R_\ast$ the flow is controlled by the stellar pulsation
(Bowen 1988), so we limit our treatment to the region 
$2 R_\ast \lesssim r \lesssim 4 R_\ast$. 
 Future numerical simulations will have to consider the regions closer to
and farther away from the star. 
 In the region $2 R_\ast \lesssim r \lesssim 4 R_\ast$ the cooling time
of the gas and dust in the shadow is very short, and we take its
temperature to be $T_{\rm sh} \simeq 300 \K$, although it can become
even cooler due to expansion.
 The average surrounding temperature is taken to be $T_e \simeq 1,200 \K$. 
 The flow time across the shaded region, i.e., in the azimuthal direction,
which we take to have a radius of $b \lesssim R_\ast$
at $r \sim 3 R_\ast$, is
$t_{\rm cross} \lesssim R_\ast/c_s \simeq 3 (R_\ast/2 \AU) \yrs$, 
where $c_s$ is the sound speed, taken at  $T_e = 1200 \K$. 
 This shows that material from a distance of $\sim R_\ast$ has time to flow
into the shadow, and by equation (2) it has time to
cool and be compressed in the shadow.  
 
 When dust forms in a high density region close to the star,
the acceleration by radiation pressure may become less efficient.
 When all the radiation is absorbed (or scattered) the momentum
imparted to the dust (and gas) by radiation is $L_\ast/c$, where
$L_\ast$ is the stellar luminosity and $c$ the speed of light.
 Let us take the mass loss rate per unit solid angle above a cool spot
to be $\dot m_{\rm sp} = \dot M_{\rm sp0} / 4 \pi$.
 If the mass loss rate per unit solid angle on the entire surface were
like that on the spot, the total mass loss rate would be $\dot M_{\rm sp0}$.
 The maximum velocity the dust and gas can reach due to radiation pressure
is
\begin{equation}
v_{\rm max} = 10 
\left( \frac{\dot M_{\rm sp0}}{10^{-5} M_\odot \yr^{-1}} \right)^{-1}
\left( \frac{L_\ast}{5,000 L_\odot} \right)
\km \s^{-1}. 
\end{equation}
 The stellar pulsations impart some momentum to the gas as well,
but much below that of radiation pressure on dust. 
 Therefore, when large quantities of dust form, the material above a cool
spot will expand at a slower velocity. 
 The material from the surroundings already has a large velocity at
$r \sim 3-4 R_\ast$ (Bowen 1988; Fleischer {\it et al.} 1992), and hence
has large velocity when entering the shaded region.
 At $r \gtrsim 4 R_\ast$ a large fraction, even most, of the material
in the shaded region is matter that has been in the surroundings,
and more material is flowing into the shadow. 
 However, since the dust blocks most of the radiation above the spot,
no further acceleration due to radiation pressure on dust occurs
in the shaded region, whereas the gravitational force of the star
decelerates the material.
 Therefore, at $r \gtrsim 10 R_\ast$, the material in the shaded
region will expand at a velocity much slower than the rest of the wind.
 The flow time of the slowly expanding gas to a distance of
$\sim 10 R_\ast \sim 10-20 \AU$ is $\sim 10~$years.
 During that time a small cool magnetic spot is likely to be destroyed,
or rotate with the star if the star rotates faster than $\sim 10^{-3}$ times
the Keplerian velocity (e.g., VY CMa; Monnier {\it et al.} 1998),
hence radiation will reach the material and accelerate it.
 This material will escape the star, but at a somewhat slower velocity
than the velocity of the surroundings.
 The lifetime of cool spots on the sun is of the order of the rotation
period, i.e., from several percents of the rotation period to a few
periods.
 The largest spots last more that the orbital period. 
 In slowly rotating AGB stars the turnover time of large convection
cells or the pulsation period are likely to determine the time scale.
 We therefore assume that small spots last no more than a few pulsational
periods, which amount to several years. 
 Fast rotating AGB stars (and other red giants like VY CMa) may support
larger cool spots which may last longer, and support enhanced mass loss
rate even in the optically-thin flow.

 At $r \sim 3-4 R_\ast$ pressure equilibrium between the shaded material
and environment is achieved for a density contrast of $\sim 4$.
 This contrast may increase a little as the wind in the shaded region
diverges less than the surrounding flow (see $\S 2$).
 Farther away from the star, the density contrast will decrease due to
the expansion of the dense material in the shaded region.
 During this phase the spot does not exist any more, and
actually there is no shadow.  
  If many small spots exist in the equatorial plane, as is expected for
magnetic cool spots, then away from the star the different high density
shaded regions will merge to form a higher density in the equatorial
plane, due both to a higher mass loss rate and the
somewhat slower expansion velocity.
{{{  This process, we suggest, can lead to an equatorial to polar density
ratio of up to $\sim 3$, hence leading to the formation of an elliptical
(but not bipolar) PN. }}}

{{{  The goal of the preliminary study in the last two sections 
is to point to some effects on the mass loss geometry which result 
from radiation shielding by dust forming close to the stellar surface. 
 Future numerical simulations of dust formation and growth and
of the flow structure above a cool spot will have to consider
several additional processes, too complicated to be studied here.
 These include the three stage cycle after a passage of a shock wave
excited by the stellar pulsation (Woitke {\it et al.} 1996),
including also the effect of the magnetic field above the magnetic cool
spot (Soker \& Clayton 1999).
 The flow of the surrounding gas into the shaded region, as discussed above,
must also be incorporated.
 Another mechanism which might make the mechanism proposed in this section
more efficient (N. Mastrodemos, private communication) is the
back-warming of the gas, which leads to periodic dust
formation not connected to the stellar pulsations
(e.g., H\"ofner {\it et al.} 1995; Fleischer {\it et al.} 1995).
}}}
   
% ======================================================================
\section{EQUATORIAL BACK-FLOW IN CLOSE BINARY STARS}
% ======================================================================

 In this section we suggest an alternative explanation for the existence
of high density equatorial material around close binary systems, where one
of the stars (the primary) is on the AGB or post-AGB phase.
 The equatorial ``disks'' extend to several$\times 100 \AU$
(Jura \& Kahane 1999), and they expand at very slow velocities.
  These systems are discussed in detail by, e.g., Waters {\it et al.}
(1993), Van Winckel, Waelkens, \& Waters (1995),
and  Van Winckel (1999, and more references therein).
 The popular explanation is that the dense equatorial material resides
in a gravitationally bound disk (e.g., Jura, Balm, \& Kahane 1995;
Van Winckel {\it et al.} 1998).
 One of the arguments used to support the rotating disk model is the
density profile of $\rho \propto r^{-1.0}$.
 However, as discussed in $\S 2$ for a cool flow, the pressure of the
surrounding hot gas results in a flow within a shaded region which
disperses less rapidly than $r^{-2}$.
 In particular, the density of an equatorial flow with constant
expansion velocity is $\rho \propto r^{-1}$. 
 Another argument, for the Red Rectangle, is the narrow spike
(full-width at half maximum of $\sim 2 \km \s^{-1}$), which may suggest
a Keplerian disk.
 Such a slow velocity is below the escape velocity, supporting
a bound disk (M. Jura, private communication).
 We note though, that the Keplerian velocity at a specific radius is
70 per cents of the escape velocity.
 Therefore, the uncertainties in the measured velocities and the distance to
the Red Rectangle do not allow to state with confidence that the velocity
is compatible with a Keplerian disk but not with a slowly expanding
equatorial flow.
{{{  We should point out that the Red Rectangle may not be a
representative of the class of systems discussed here (H. Van Winckel,
private communication), because of its C-rich extended nebula and O-rich
circumbinary disk. }}}
 
 We find a problem in maintaining Keplerian disks at radii larger
by two orders of magnitude than the binary orbital separation.
The disk in the Red Rectangle, for example, has an outer radius of
several$\times 100 \AU$ (Waters {\it et al.} 1998; Jura, Turner, \&
Balm 1997).
  The angular momentum per unit mass is more than an order of magnitude
larger than that of the binary system itself. 
 If a tertiary star exist (Jura \& Turner 1998), it may supply the required
angular momentum in the Red Rectangle (M. Jura, private communication).
 The problem still remains with the other systems. 
 Mastrodemos \& Morris (1999, hereafter MM99) find that in interacting
binaries with orbital periods similar to those of these systems,
$\sim 100-2000$ days (Van Winckel 1999), most of the mass leaves in
the equatorial plane.
 This equatorial flow will interact with any equatorial disk, and will
destroy it, unless the disk is very massive, i.e., has survived from the
pre-main sequence phase.
However, the chemistry of these disks suggests that the material is from 
AGB winds.
 To maintain the disk, the material in the equatorial flow must have
a specific angular momentum similar to that of the disk.
  This seems to be too much angular momentum to be supplied by the binary
system, as the total mass lost is $\gg 10 \%$ of the binary system mass.
 Instead, we suggest a slowly expanding, and even back flowing, equatorial
dense flow, with a very small amount of angular momentum, more than an
order of magnitude below that required by a disk.

 We suggest that because of radiation shielding, the radiation
pressure on dust and gas in the equatorial plane away from the binary
system is not efficient.
 Hence, the expansion velocity is very slow $\sim 1 \km \s^{-1}$.
 This slow expansion still allows for dust to be decoupled from gas,
and allows the peculiar abundance found in these systems, although no
Keplerian disk is formed.
 Mastrodemos \& Morris (1998; 1999) show that in close binary systems
where the secondary is not much lighter than the primary star,
most of the wind is diverted toward the equatorial plane.
 The simulations of MM99 show that the
expansion velocity in the equatorial plane is somewhat slower
that that above and below it, but in all their simulated cases the
matter in the equatorial plane reaches the escape velocity from the system.
Mastrodemos \& Morris, though, did not consider any shielding by the
dust close to the binary system.
  According to the theme of the present paper, if mass loss rate is high
enough, radiation is blocked in the equatorial direction and the material
is not accelerated any more.
 Because of gravitational torque by the secondary, the equatorial
material has a large azimuthal velocity and hence it will expand to
large distances from the system.
 For example, in their model 1, MM99 find
an azimuthal velocity of $\sim 15 \km \s^{-1}$ at a little more
than the orbital separation of $3.6 \AU$.
 The mass loss rate per steradian in the equatorial plane in Mastrodemos
\& Morris's model 2 is 
$\dot m_{eq} \simeq 7.5 \times 10^{-5}/(4 \pi) M_\odot \yr^{-1}$.
  This equatorial mass loss is achieved for a total mass loss rate of
$\sim 2 \times 10^{-5} M_\odot \yr^{-1}$.
 An even higher mass loss rate in the equatorial flow will be attained 
by a higher mass loss rate from the AGB star.
 From equation (2), radiation from a typical AGB star cannot support
such a flow. 
 The radiation from the star can blow a more spherical 
wind, with a total mass loss rate of 
$\dot M \sim$few$\times 10^{-5} M_\odot \yr^{-1}$. 
 However, the close secondary diverts the flow toward the equatorial plane,
where the mass loss rate per unit solid angle is much higher. 
 Therefore, the flow will reach a very low terminal velocity. 
When the flow velocity is below the drift velocity between dust grains
and the gas particles, the gas may flow back toward the star.
 
 To demonstrate this behavior we assume the following simple flow
structure, based on the results of MM99.
 They find that in close binaries most of the mass
leaves the system close to the equatorial plane in a spiral equatorial flow.
 The density near the secondary is very high, so we argue
that when mass loss rate is high enough and concentrated near the
equatorial plane, the radiation will be blocked by the first,  
innermost spiral loop.
 A possible collimated outflow from the secondary (mass accreting) star
may further concentrate the outflow  toward the equatorial plane
(Soker \& Rappaport 2000). 
 Therefore, the matter is accelerated by radiation only during one
orbital period, when it still belongs to the inner loop.
 A support to this assumption regarding the spiral structure comes
from a different kind of binary system, the WR-OB binary system WR104.
 A dusty outflow in a one-coil structure extending to a distance
of $\sim 150 \AU$ from the binary system is observed in this
system (Tuthill, Monnier, \& Danchi 1999).
 They suggest that the dusty one-coil outflow is the innermost loop of a
more extended spiral structure.
 The innermost loop obscures the other coils from the radiation
of the stars, and hence no IR radiation is detected from them.
 The binary system is composed of a WR star and an OB star, with an orbital
period of 220 days, as inferred from the rotation of the spiral
structure which was observed at two epoch.
 There are several differences between this system and the systems studied
by MM99 and in the present paper; the most significant are.
($i$) The dust in this type of binary systems does not form near the
surface of the mass-losing star, but where the two winds collide.
($ii$) The wind velocity is much larger than the orbital velocity, and
reaches its terminal velocity at a radial distance from the binary system
much smaller than the distance of the innermost loop outer point
at $\sim 150 \AU$.

 Let the mass loss rate per unit solid angle in the equatorial flow be
$\dot m_{eq}=\dot M_{eq} /4 \pi$, $M_1$ and $M_2$ are the masses of the
primary (mass losing) and secondary stars, $a$ is the orbital separation,
$P_{\rm orb}=2 \pi [G(M_1+M_2)/a^3]^{-1/2}$ is the orbital period, and
$L_\ast$ is the primary luminosity.
 We also assume that most of the wind resides in the spiral structure,
that the optical depth of the inner spiral loop is $>>1$, and that along
radial directions above and below the equatorial plane the optical depth
is $<<1$.
 The mass per unit solid angle along a segment of the spiral structure is
$\Delta m = \dot m_{eq} P_{\rm orb}=\dot M_{eq} P_{\rm orb}/4 \pi$.
 If the optical depth is large, from momentum conservation we
find the radiation acceleration to be
$(d^2r/dt^2)_{\rm rad} = L_\ast /(4 \pi c \Delta m)$.
 The equation of motion for a parcel of gas includes gravity by the
binary system and the centrifugal force.
 We assume axisymmetry around the primary star. 
 This assumption does not hold close to the secondary, where the spiral
structure starts (MM99), and the spiral 
structure is around the center of mass rather than the primary. 
However, when the primary is more massive, these assumptions are adequate
for the present simple treatment. 

 The equation of motion for a parcel of gas in the spiral flow, and 
for a time of one orbital period, reads
\begin{equation}
\frac{d^2r}{dt^2} = 
\frac {L_\ast}{c \dot M_{eq} P_{\rm orb}}
+\frac{v_t^2}{r} -\frac {G(M_1+M_2)}{r^2}.
\end{equation}
  From angular momentum conservation we take the azimuthal velocity to 
be $v_t =v_{t0} a/r$, where $v_{t0}$ is the azimuthal velocity where the
flow starts, at $r=a$. We parameterize it by 
$v_{t0}= \beta (2 \pi a/P_{\rm orb})$. 
 We scale variables with a length unit $a$ and a time unit
$P_{\rm orb}/2 \pi$, and define the dimensionless variables
$x \equiv r/a$ and $\tau \equiv 2 \pi t/P_{\rm orb}$. 
 Using the expression for $P_{\rm orb}$ we write equation (5) as
\begin{equation}
\frac{d^2x}{d \tau ^2} = 
\eta+\beta^2 x^{-3} - x^{-2}, 
\end{equation}
where we defined 
\begin{equation}
\eta= \frac {L_\ast P_{\rm orb}}{4 \pi^2 c a \dot M_{eq}} 
=0.076 
\left( \frac{L_\ast}{10^4 L_\odot} \right)
\left( \frac{\dot M_{eq}}{10^{-5} M_\odot \yr^{-1}} \right)^{-1}
\left( \frac{a}{1 \AU} \right)^{1/2}
\left( \frac{M_1+M_2}{2 M_\odot} \right)^{-1/2}.
\end{equation}
 We can analytically integrate equation (6) once.
 With the initial conditions of $dx/ d\tau = \alpha$ at $x=1$,
the integration gives 
\begin{equation}
v_x = \frac{dx}{d \tau } = 
[ 2 \eta x -\beta^2 x^{-2} + 2 x^{-1} + \alpha^2 - 2 \eta +\beta^2 -2 ]^{1/2}.
\end{equation}
 When $\alpha=1$ and $\beta=1$ the matter leaves the binary at 
the escape velocity from $x=1$. 
 We stress again that the above solution is adequate when the mass in
the inner spiral loop in the equatorial plane (and only in the
equatorial plane) absorbs or scatters all the incident radiation,
hence it should be integrated for only one orbital period,
from $\tau=0$ to $\tau=2 \pi$.

 We integrate equation (8) over one orbital period, and find the location
$x_f$ of a parcel of gas on the spiral structure at that time, and the ratio
of its  velocity to the escape velocity at that location
$w \equiv (v_r^2+v_t^2)^{1/2}/v_{\rm esc}$,
{{{ where $v_r$ is the radial velocity. }}}
 In Figure 2 we plot $w$ (solid lines) and $x_f$ (dashed lines) as a
function of $\eta$ for $\beta=1$ and three values of $\alpha$,
as indicated on each line.
 The value of $\beta=1$ indicates strong interaction with the companion
star, so this case is applicable to a close orbital separation.
  At this distance the wind does not yet reach the escape velocity at
the orbital separation,
{{{ which has a value of $\alpha=2^{1/2}$ in our dimensionless units. }}}
 Therefore, for these cases we expect $\alpha \ll 2^{1/2}$.
 The solid lines (velocity divided by escape velocity at $x_f$)
are plotted only for cases where the flow has a positive radial velocity.
 When the flow stagnates in the radial direction, i.e., $v_x=0$ at
$\tau < 2 \pi$, the dashed lines indicate the location of the stagnation
distance (in orbital separation units).
  For larger orbital separations the wind is faster and the secondary
influences the wind less, so the azimuthal velocity will be lower.
 In Figure 3 we plot $w$ (solid lines) and $x_f$ (dashed lines) as a
function of $\eta$ for $\alpha=1$ and three values of $\beta$,
as indicated on the graph.
  The assumptions that led to equation (8) and Figures 2 and 3, break down
for large values of $\eta$ and/or large values of $x_f$, which means low
mass loss rates, since then the flow is not optically thick and not
all radiation is absorbed.
 Therefore, the results for $w \gtrsim 5$ are not accurate.
 When radiation pressure dominates and $\beta=1$, so that the centrifugal
force balances the gravity near $x=1$, the following approximation holds
$x_f \sim 2 \pi \alpha + \eta (2 \pi)^2/2$. 
 This explains the linear relation between $x_f$ and $\eta$ with a slope of
$\sim 2 \pi^2$, for large values of $x_f$ when $\beta=1$.
  In any case, the interesting region of the graph is where the
escape velocity is $w \sim 1$. 

  In the models of MM99, the first spiral
loop reaches a distance of $x_f \sim 3-8$.
  For similar ranges of $x_f$, we see from Figures 2 and 3 that the
material reaches the escape velocity only for $\eta \gtrsim 0.05$.
  The flow cannot have $\eta \ll 0.1$, since this means too high
a (spherical) mass loss rate from the mass-losing star, which cannot
be supported by its radiation.
 Therefore, when the wind from the mass-losing star is on the limit
of what its radiation can support, and a secondary star concentrates
the flow toward the equatorial plane, a dense, slowly expanding
equatorial flow will result.
 For that to happen the conditions are $\eta \lesssim 0.1-0.2$ and an
optically thick flow.
 The flow will not be accelerated any more when away from the star,
because of an optically thick equatorial flow, and it will be slowed
down by the gravity of the binary stars.

  The primary (mass-losing) stars of the binary systems discussed at the
beginning of this section are depleted of refractory elements, which
compose the dust particles (e.g., Van Winckel {\it et al.} 1998).
  In a mechanism proposed by Waters, Trams \& Waelkens (1992) the
gas decouples from the dust in a relatively low density.
 While the dust particles are accelerated by radiation pressure, some of
the gas is accreted back to the star.
 For this mechanism to be efficient, the gas and dust should reside in
a long-lived disk.
 In the mechanism proposed here, the separation of dust from gas
occurs above and below the equatorial flow, where material comes
from the slowly expanding equatorial flow.
 Because of the slow radial expansion, material spends a long
time there, allowing the decoupling of dust from gas to be efficient. 
 In these regions, above and below the dense equatorial flow, the
density is relatively low, and radiation can accelerate the dust
particles, unlike the equatorial plane, where the density is higher
and not much radiation penetrates.
 The grains above and below the disk absorb and re-radiate in the infrared.
 This radiation may accelerate the material in the disk, so that even if
the disk material outflows at a velocity somewhat below the escape
velocity, it will be further accelerated.
 In any case, the flow will still be slow.

 We claim that the slowly expanding equatorial flow model
{{{ may }}} explain the six arguments listed by Jura \& Kahane
(1999) to support orbiting material, while having the advantage that
it does not require a huge amount of angular momentum.
 {{{ Their six arguments can by summarized in short as follows:
 (1) Narrow CO line profile. (2) The abundance of the central star
HD 44179 can be explained by accretion of surrounding gas but not dust.
 (3) Presence of compact massive disk. 
 (4) Presence of Grains as large as $0.02 \cm$.
 (5) Presence of crystalline oxygen-rich silicate, which require
long-lived flow or disk to form. 
 (6) The eccentricity of the orbit of the binary system can
 be explained by the presence of a long-lived circumbinary disk.

 The presently proposed model also has a slowly moving material. }}}
 The slow expansion means that the material spends a long
time in the radiation field of the star; it takes 250 years for
material moving at $2 \km \s^{-1}$ to cross $100 \AU$.
 Gas that moves even slower, part of which is accreted back on the star,
will spend a much longer time in these regions.
 This time scale of $\sim 10^3 \yrs$, is a substantial fraction of
the lifetime of post-AGB stars.
 The formation of large grains occurs close to the star after the
dust has become shielded and as a result cools quite fast. 
{{{  The high eccentricity can be explained by a higher mass
loss rate at periastron passage (e.g., Harpaz, Rappaport, \& Soker 1997),
rather than a massive disk. }}}

{{{ There are two processes that should be explored in more detailed
by future calculations (H. Van Winckel, private communication).
 Firstly, the systems mentioned in this section have typical orbital
periods of $\lesssim 1 \yr$, shorter than the orbital periods of
the systems studied by MM99.
 The mass loss from these closer binary systems should be followed
by 3D numerical simulations to find whether a spiral structure is
formed.
 Secondly, an argument used to support the long lived disk model is
the presence of crystalline dust together with large grain sizes.
 The possibility of dust crystallisation and formation of large
grains in the proposed alternative scenario of dense and cold equatorial
flow should be explored. }}}

% ======================================================================
\section{DISCUSSION AND SUMMARY}
% ======================================================================
 
 The main goal of this paper was to strengthen the {\it cool magnetic spots}
model, by proposing a process which further increases the inhomogeneity 
of the wind from stars which are about to leave the AGB. 
 This model was developed to explain the increase in the departure from
sphericity of the mass lost from very slowly rotating stars as they
evolve on the upper AGB, and early post-AGB phase.  
 Such an increase is inferred from the larger departure from spherical
structure of the inner regions of many elliptical PNs.  
 Most of these elliptical PNs have no close stellar companions, but
evolve as single stars or are spun-up by planets, and hence
are very slow rotators when reaching the upper AGB
(Soker \& Harpaz 1999 and references therein).
  The cool magnetic spots model is based on the assumption that a very
slow rotation of an AGB star with very low mass envelope is sufficient
to maintain a weak magnetic activity, which forms cool magnetic
spots near the equatorial plane (Soker 1998; Soker \& Harpaz 1999).
  Large quantities of dust are expected to be formed above the surface of
these cool spots (Soker \& Clayton 1999).

  We proposed here that dust which is formed very close to the surface
of a cool spot, practically at its surface, during a high mass loss rate
phase, has a large optical depth, and it shields the region above it from
the stellar radiation.
 As a result the temperature in the shaded region decreases rapidly
relative to the surrounding temperature.
 This leads to both further dust formation in the shaded region
and, due to lower pressure, the convergence of the stream toward
the symmetry axis of the conical shaded region (see fig. 1).
 This process increases the departure from sphericity in two ways.
 First, the shielding by dust allows dust formation above small cool
spots.
 Without the formation of dust close to the surface of the spot and the
shielding, only large spots, with radii $b_s \gtrsim 0.3 R_\ast$,
allow enhanced dust formation (Frank 1995).
 Second, because of the converging flow, the density in the shaded region
away from the star, up to $\sim 100 \AU$, decreases less steeply than
$r^{-2}$.
 Hence, the density contrast between the shaded region and
the environment increases.
  In the present preliminary study we estimate that the initial density
enhancement at $r \sim 5 R_\ast$ can be up to a factor of $\sim 4$. 
{{{  Future numerical calculations should examine
whether the density enhancement }}}
can further increase away from the star.

  In previous papers (Soker 1998; Soker \& Harpaz 1999) the positive
correlation between high mass loss rate (in particular the superwind phase
on the termination of the AGB) and the larger departure from sphericity
was attributed to a more effective magnetic activity.
 This activity was assumed to form  larger and more numerous cool
magnetic spots.
 The process discussed in the present paper further increases the
correlation since it is effective for small spots, but only when
mass loss rate is high, as in the superwind phase, hence optical
depth is large.
  When the flow is optically thin, large cool spots are required to
form significant higher mass loss rate (Frank 1995).
 
 The local enhanced dust formation and formation of a concentrated flow
lead to the formation of filaments, loops, and arcs in the circumstellar
medium and later in the PN.
 Many PNs show such structures, which may be formed by this process,
or from instabilities later in the evolution.
 Concentration of many cool spots near the equatorial plane will lead
in addition to the formation of an elliptical PN.

 Close binary systems where the secondary star diverts most of the mass
lost by the primary toward the equatorial plane, and the mass loss occurs
in a spiral structure (MM99), are another type
of systems in which shielding by dust may influence the mass loss process.
 If both mass loss rate and the concentration toward the equatorial plane
are large, then the innermost loop of the spiral structure shields the rest
of the equatorial plane from the stellar radiation.
 This prevents the material farther out from acceleration by radiation
pressure, and a dense, slowly expanding, equatorial flow is formed.
 Dense equatorial matter is found at several hundred astronomical units
around several close binary systems, where one of the stars is a post-AGB
star (or a similar object).
 We suggest that such a slowly expanding equatorial flow can explain
the structure of the circumstellar matter around these systems.
 This type of flow model has the advantage that it does not require
large angular momentum, unlike the model of a gravitationally bound
disk, which requires too large an angular momentum to be supplied
by the binary system.
 
{\bf ACKNOWLEDGMENTS:} 
{{{ I thank the referee, Nikos Mastrodemos, for very detailed and
helpful suggestions, }}}
and Geoff Clayton, Mike Jura
{{{ and  Hans Van Winckel }}} for helpful comments.
 This research was supported in part by a grant from
the Israel Science Foundation.

%\newpage

\centerline {\bf FIGURE CAPTIONS}

{\noindent {\bf Figure 1:} A schematic drawing of the flow structure above
a cool spot.  
The shaded region represents the density of dust (upper panel), 
while arrows indicate the flow direction (but not its speed; lower panel).
(ask the author for the figure). 

{\noindent {\bf Figure 2:} The location and velocity of a parcel of
gas after one binary orbital period $P_{\rm orb}$.
 The flow is assumed to be in an equatorial spiral structure, and
optically thick.
 The location (dashed line) is given in units of the orbital separation,
and the velocity (solid line) in units of the escape velocity at that
location.
 The plots are for $\beta=1$, and three values of $\alpha$, as indicated,
where $\alpha$ and $\beta$ are the radial and azimuthal velocity of the
flow when it leaves the binary system, respectively, both in units of
$2 \pi a /P_{\rm orb}$.

{\noindent {\bf Figure 3:} Like figure 2, but for $\alpha=1$ and three
values of $\beta$, as indicated on the graph.

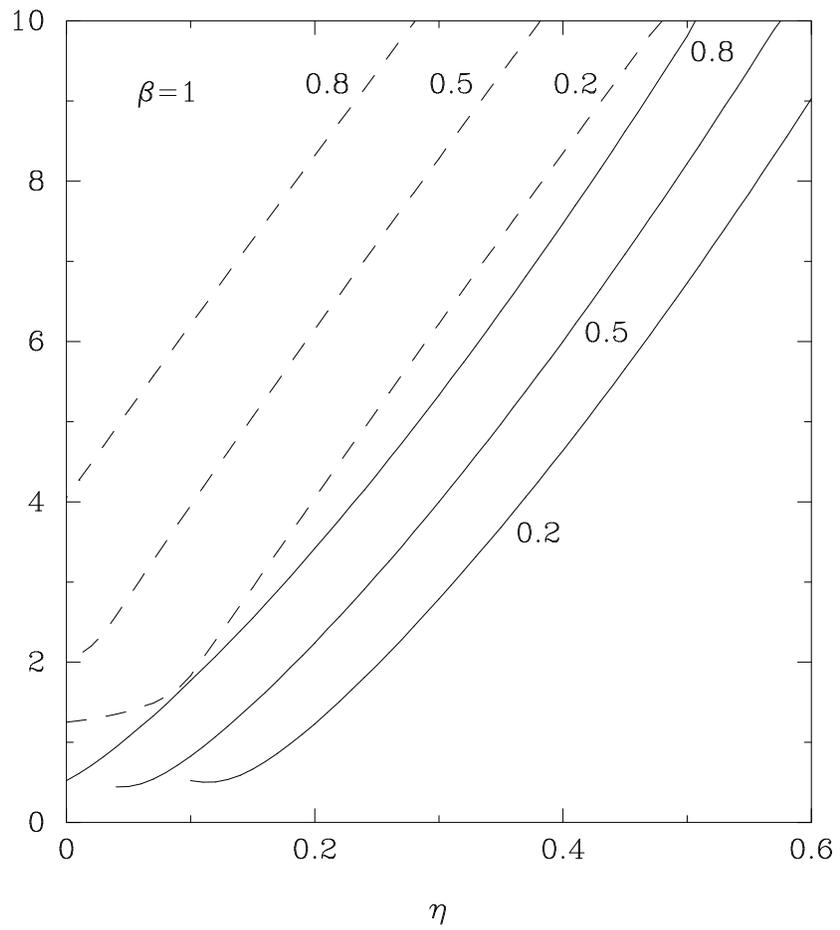
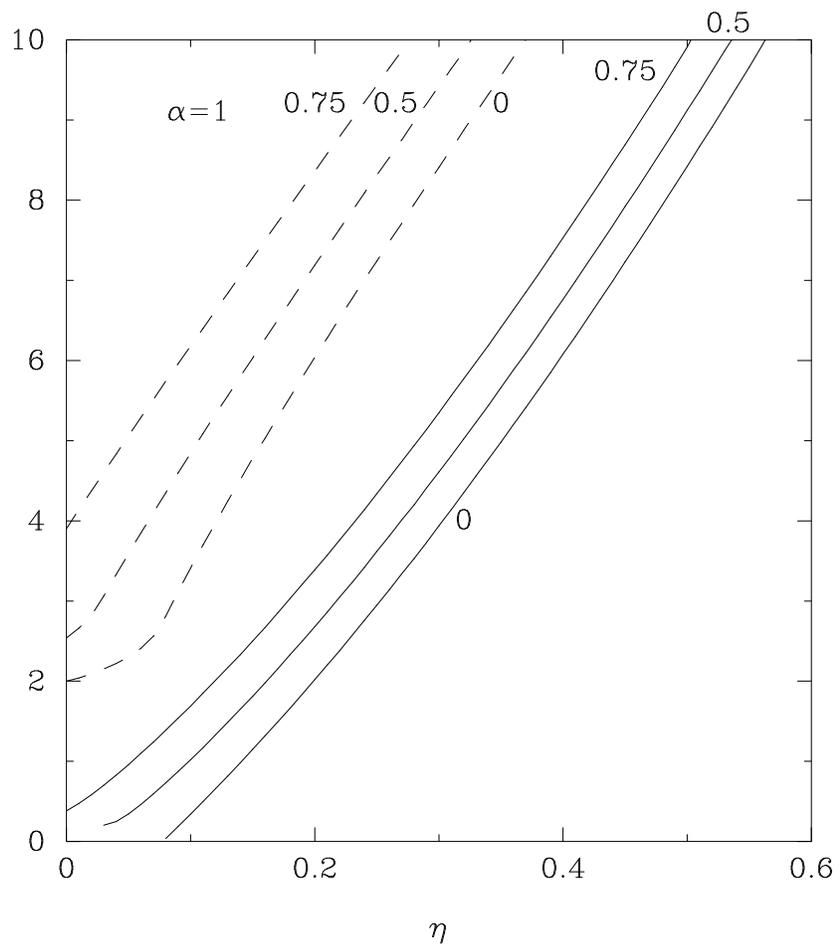
\end{document}